\documentclass[12pt]{article}
\usepackage{graphicx}
\usepackage{cite}
\newcommand\pubnumber{}
\newcommand\pubdate{\today}

\def\institute{Joint Laboratory of Optics of Palacký University Olomouc and Institute of Physics of Czech Academy of Sciences, Czech Republic}

\def\authemail{\footnote{Contact: petr.baron@upol.cz\\[12pt]“Copyright 2023 CERN for the benefit of the ATLAS Collaboration. Reproduction of this article or parts of it is allowed as specified in the CC-BY-4.0 license.”}}

\def\Title#1{\begin{center} {\Large #1 } \end{center}}
\def\Author#1{\begin{center}{ \sc #1} \end{center}}
\def\Address#1{\begin{center}{ \it #1} \end{center}}

\newcommand\pubblock{\rightline{\begin{tabular}{l} \pubnumber\\
         \pubdate  \end{tabular}}}
\newenvironment{Abstract}{\begin{quotation}  }{\end{quotation}}
\newenvironment{Presented}{\begin{quotation} \begin{center} 
             PRESENTED AT\end{center}\bigskip 
      \begin{center}\begin{large}}{\end{large}\end{center} \end{quotation}}

\usepackage[font=small,labelfont=bf]{caption} % Required 
\usepackage{url}            % simple URL typesetting
\usepackage{eprint}

\begin{document}
\begin{titlepage}
\pubblock

\vfill
%\Title{Observation of $t\bar{t}$ production in lepton+jets and dilepton channels in $p$+Pb collisions at $\sqrt{s_\mathrm{NN}}=8.16$~TeV with the ATLAS detector}
\Title{Observation of top-quark pair production in proton-lead collisions in ATLAS}
\vfill
\Author{ Petr Baroň\authemail on behalf of the ATLAS Collaboration}
\Address{\institute}
\vfill
\begin{Abstract}
Top-quarks and Higgs boson are the only elementary particles that have not been observed in heavy-ion collisions in the ATLAS detector yet. In particular top quarks, the heaviest elementary particles carrying colour charges, have been argued to be attractive candidates for probing the quark-gluon plasma produced in heavy-ion collisions. In proton-lead collisions, top-quark production is expected to be sensitive to nuclear modifications of parton distribution functions (PDF) at high Bjoerken-x values which are hard to access experimentally using other probes available so far. In 2016 the ATLAS experiment collected proton-lead collisions at centre-of-mass energy of 8.16 TeV per nucleon pair. The data sample corresponds to an integrated luminosity of 164 nb$^{-1}$, which allows for the first time in this data set with ATLAS, to measure top-quark pair production. In this work, we discuss the inclusive cross section measurement for the top-quark pairs production in dilepton and lepton+jets decay modes with electrons and muons recorded by the ATLAS experiment. The measurement is compared to the NNLO predictions for top-quark production using various PDF sets.
\end{Abstract}
\vfill
\begin{Presented}
$16^\mathrm{th}$ International Workshop on Top Quark Physics\\
(Top2023), 24--29 September, 2023
\end{Presented}
\vfill
\end{titlepage}
\def\thefootnote{\fnsymbol{footnote}}
\setcounter{footnote}{0}

\section{Introduction}
This proceedings showcases the process of $t\bar{t}$ production in \pPb{} collisions at $\sqrt{s_\mathrm{NN}}=8.16~\mathrm{TeV}$ measured by the ATLAS experiment. 
A comprehensive exposition is available in the recent conference note \cite{confnote}. 
The study focuses on the reconstruction of top-quark pairs observed in both \ljets{} and dilepton channels, 
employing electrons and muons in the final states. While the dilepton mode is less abundant, its purity significantly surpasses that of the \ljets{} channels. The results are compared with NNLO calculations involving the most recent nPDF sets.\par

The ATLAS detector \cite{ATLAS:2008xda,Capeans:1291633,ATLASIBL:2018gqd,ATLAS:2016wtr,ATLAS:2020esi,ATL-SOFT-PUB-2021-001,ATLAS:2019fst} collected the data for this measurement during the 2016 \pPb{} collision running period,
 amassing an integrated luminosity of 165~nb$^{-1}$. Operation of the LHC at energies of 6.5~TeV for the proton beams and 2.51~TeV per nucleon for the Pb beams led 
 in a center-of-mass collision energy of 8.16~TeV. This configuration induces a rapidity boost of $\pm0.465$ units w.r.t. the ATLAS laboratory frame, 
 depending upon the direction of the $p$ beam.\par

The experiment facilitated two beam-direction configurations: \pPb{} and \Pbp{}. The latter configuration, with the Pb beam directed in the $+z$ direction, 
provided approximately twice the integrated luminosity compared to the former.

%Event reco and selection
\section{Data and Monte Carlo Simulation Samples}

The ATLAS experiment used data collected during the 2016 \pPb{} running period, amounting to an integrated luminosity of 165~nb$^{-1}$. 
Events were selected via single-lepton electron or muon triggers with a minimum transverse momentum $p_{\mathrm{T}}$ threshold of 15~GeV 
\cite{ATLAS:2019dpa,ATLAS:2020gty}. These events were required to possess at least one reconstructed vertex built from a minimum of two 
good-quality charged tracks with $p_{\mathrm{T}} > 0.1~\mathrm{GeV}$. Electron candidates were identified through a cluster of energy deposits in the EM calorimeter matched to an ID track. 
Selection criteria included 'Medium' likelihood-based requirements \cite{ATLAS:2019qmc}, $p_{\mathrm{T}} > 18~\mathrm{GeV}$, and $|\eta| < 2.47$. 
Varying isolation requirements were imposed based on track and calorimeter-based isolation, with efficiency thresholds at different $p_{\mathrm{T}}$ values. 
Muon candidates were reconstructed by combining inner detector (ID) and muon spectrometer (MS) tracks, satisfying 'Medium' requirements \cite{ATLAS:2020auj} with $p_{\mathrm{T}} > 18~\mathrm{GeV}$, 
$|\eta| < 2.5$, and isolation criteria using a cone size of $\Delta R=0.2$ around the muon and a track-based isolation
requirement with a varying cone size~\cite{ATLAS:2020auj}. Jets were reconstructed from calorimeter energy deposits~\cite{ATLAS:2019jgo, HION-2018-17}, using the anti-$k_t$ algorithm~\cite{Cacciari:2008gp,Cacciari:2011ma} 
with $R = 0.4$. The jet kinematics are corrected event-by-event for the contribution from 
underlying event (UE) particles, calibration included calorimeter response simulations \cite{ATLAS:2020cli}, and in situ absolute energy scale measurements. 
In situ measurements are carried out in $pp$ collisions and cross-calibrated to the $p+Pb$ system typical by its low pile-up ($<\mu> = 0.18$) enviroment. Jets reconstructed this way are referred to as Heavy Ion (HI) jets. HI jets and Particle Flow (PF) jets, were matched using a minimal $\Delta R$ criterion. HI jets lacking PF counterparts were considered non-$b$-tagged. 
The analysis relied on HI jets for kinematic calculations, but inherited $b$-tagging information from matched PF jets.

Monte Carlo (MC) simulated events were utilized for analysis development, signal estimation, and background evaluation. 
Samples were processed using the ATLAS detector simulation based on \geant4~\cite{GEANT4:2002zbu}. 
Top-quark samples were simulated using various event generators \cite{Alioli:2010xd,SJOSTRAND2008852,ATL-PHYS-PUB-2014-021}, 
normalized using NNLO+NNLL predictions \cite{Czakon:2011xx}. Backgrounds from $W/Z$+jets, single top-quark, and diboson production were simulated using \sherpa{} and \powheg{} event generators 
with \pythia8 for particle shower. The fake-lepton and non-prompt background was estimated using the Matrix Method~\cite{ATLAS:2022swp}. Events were categorized into dilepton and \ljets{} channels based on lepton count 
and $b$-tagged jet requirements, with specific event yields and signal purities observed.

The dilepton channel comprises events featuring precisely two opposite-sign leptons. Pairs of same-flavor leptons ($e^+e^-$ or $\mu^+\mu^-$) within an invariant 
mass range of $80<m_{\ell\ell}<100~\mathrm{GeV}$ are discarded. Additionally, the invariant mass in the $e\mu$ ($ee$ and $\mu\mu$) 
channel must exceed 15 (45)~GeV, a constraint aimed at aligning with the phase space of the $Z$+jets simulation samples without significantly impacting the results. 
Events with a minimum of two HI jets with at least one being $b$-tagged form the Signal Region (SR). 
In total 106 and 104 data events were observed in the $1b$ and $\geq2b$ regions, respectively, exhibiting expected signal purities of 53\% and 91\% in data.
The lepton+jets~(\ljets) channel is constructed from events featuring precisely one lepton and at least four HI jets, including at least one $b$-tagged jet, 
in the SR. This channel comprises 1874 and 1075 data events in the $1b$ and $\geq2b$ regions, respectively, exhibiting anticipated signal purities of 20\% and 71\%. 
Further division of the \ljets{} SR, based on lepton flavor, produces in total four signal regions: $1\ell1b$~ejets, $1\ell2b\mathrm{incl}$~ejets, $1\ell1b$~mujets, and $4j1\ell2b\mathrm{incl}$~mujets.

\section{Systematic uncertainties}

The measurement is subject to systematic uncertainties stemming from various sources:
\begin{itemize}
  \item Lepton reconstruction uncertainties encompass muon momentum scale and resolution based on existing references \cite{ATLAS:2020auj}. 
  For electrons, reconstruction, identification, isolation, and low-pileup energy calibration uncertainties are assessed~\cite{ATLAS:2019jvq}.
  \item Jet-related uncertainties rely on studies of calorimeter response and comparisons between different simulation samples \cite{ATLAS:2017bje,ATLAS-CONF-2015-016}. 
  These uncertainties also cover $b$-tagging and jet matching variations.
\item To account for uncertainties in $V$+jets processes, single-top-quark processes, and the diboson background, the Berends scaling technique, diagram removal/subtraction variations, 
and normalization uncertainties are utilized \cite{PhysRevLett.106.092001,Re:2010bp,ATLAS:2022jbj}.
\item Uncertainties in the non-prompt and fake-lepton background involve statistical and systematic variations in efficiencies. Shape variations in the \ljets{} channel and 
normalization uncertainties are imposed based on conservative estimations in the $0b$ control region.
\item Effects from the QCD initial-state and final-state radiation, evaluated via variations in $\alpha_{\mathrm{s}}$, renormalization/factorization scales, 
$h_\mathrm{damp}$ parameter in \powheg{}, and PDF uncertainties, are considered \cite{ATL-PHYS-PUB-2014-021,Butterworth:2015oua}.
\item The integrated luminosity uncertainty is estimated at 2.4\%. This evaluation is based on luminosity scale calibration using 
$x$-$y$ beam-separation scans and measurements from the LUCID-2 detector \cite{ATLAS:2022hro,LUCID2}.
\end{itemize}

\section{Analysis}
\begin{figure}[!htb]    
\includegraphics[width=0.32\textwidth]{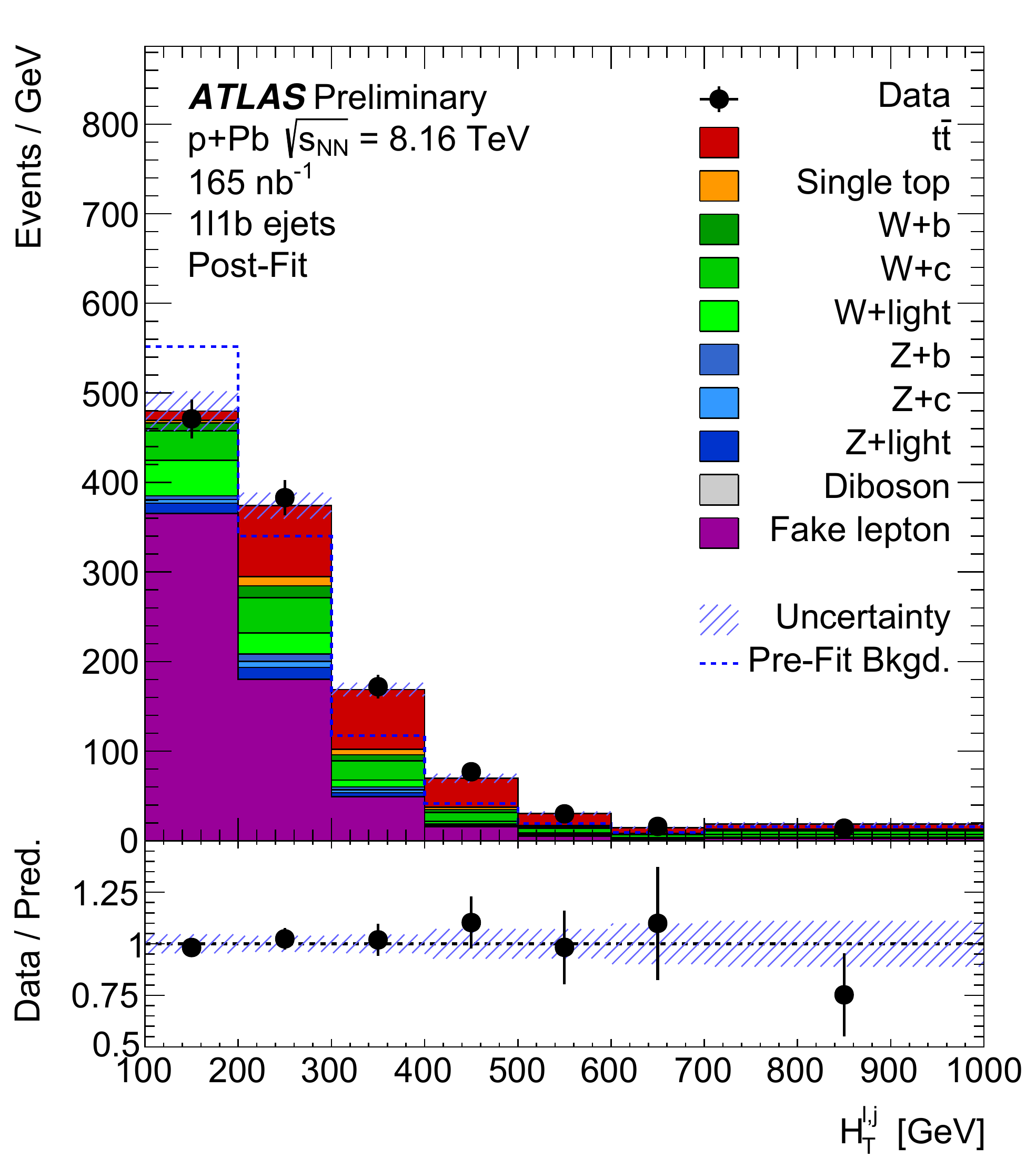}
\includegraphics[width=0.32\textwidth]{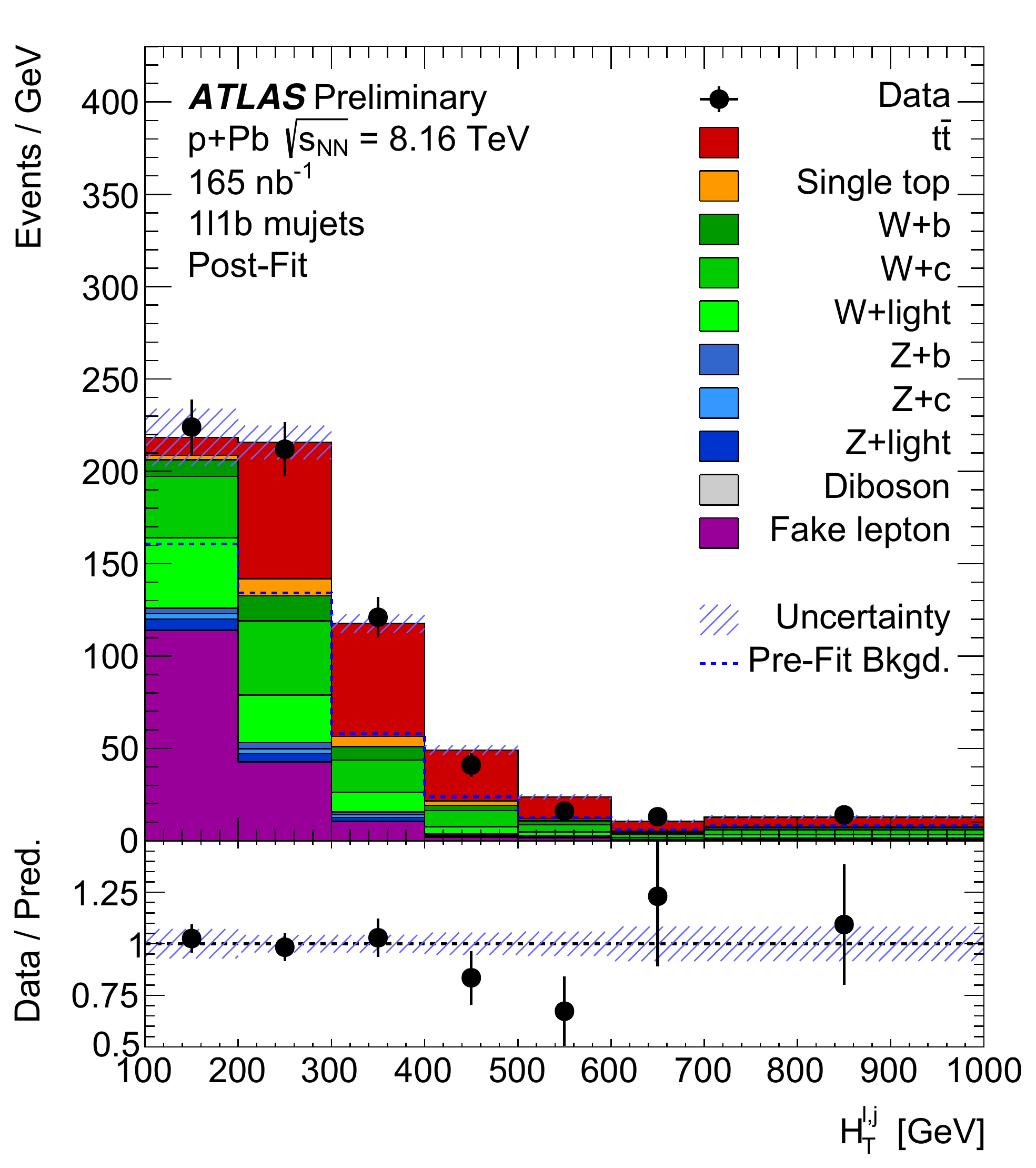}
\includegraphics[width=0.32\textwidth]{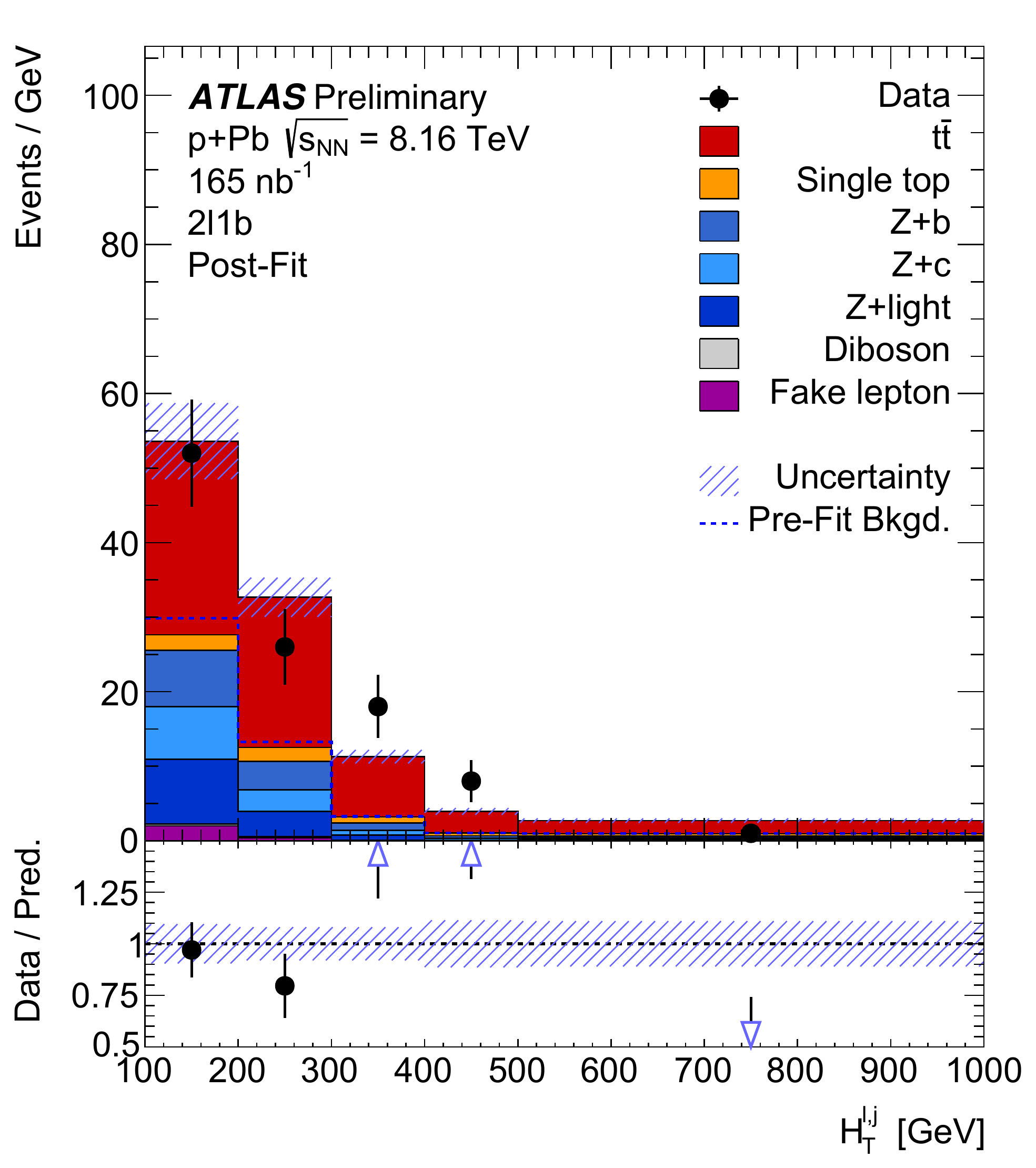} \\
\includegraphics[width=0.32\textwidth]{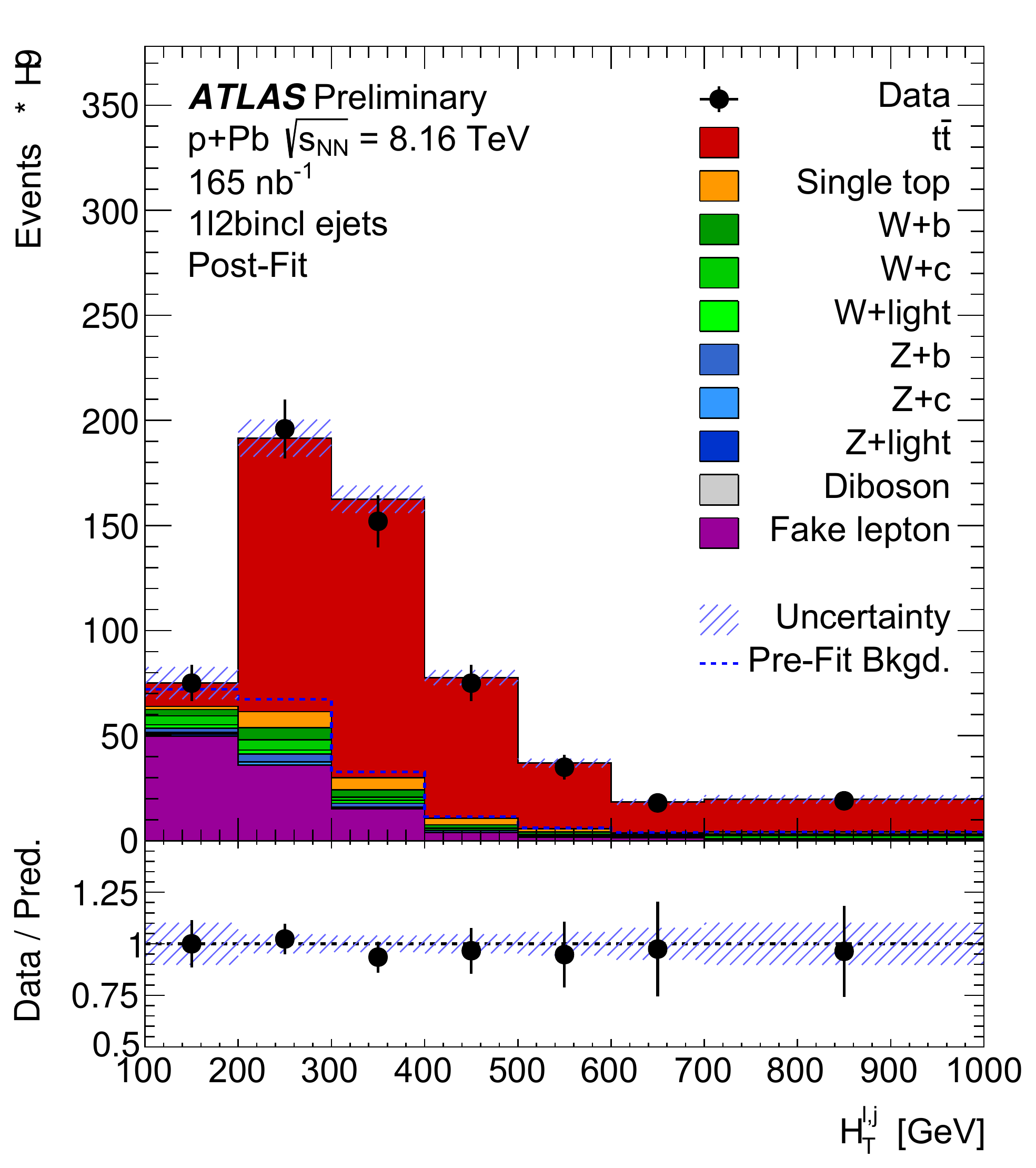}
\includegraphics[width=0.32\textwidth]{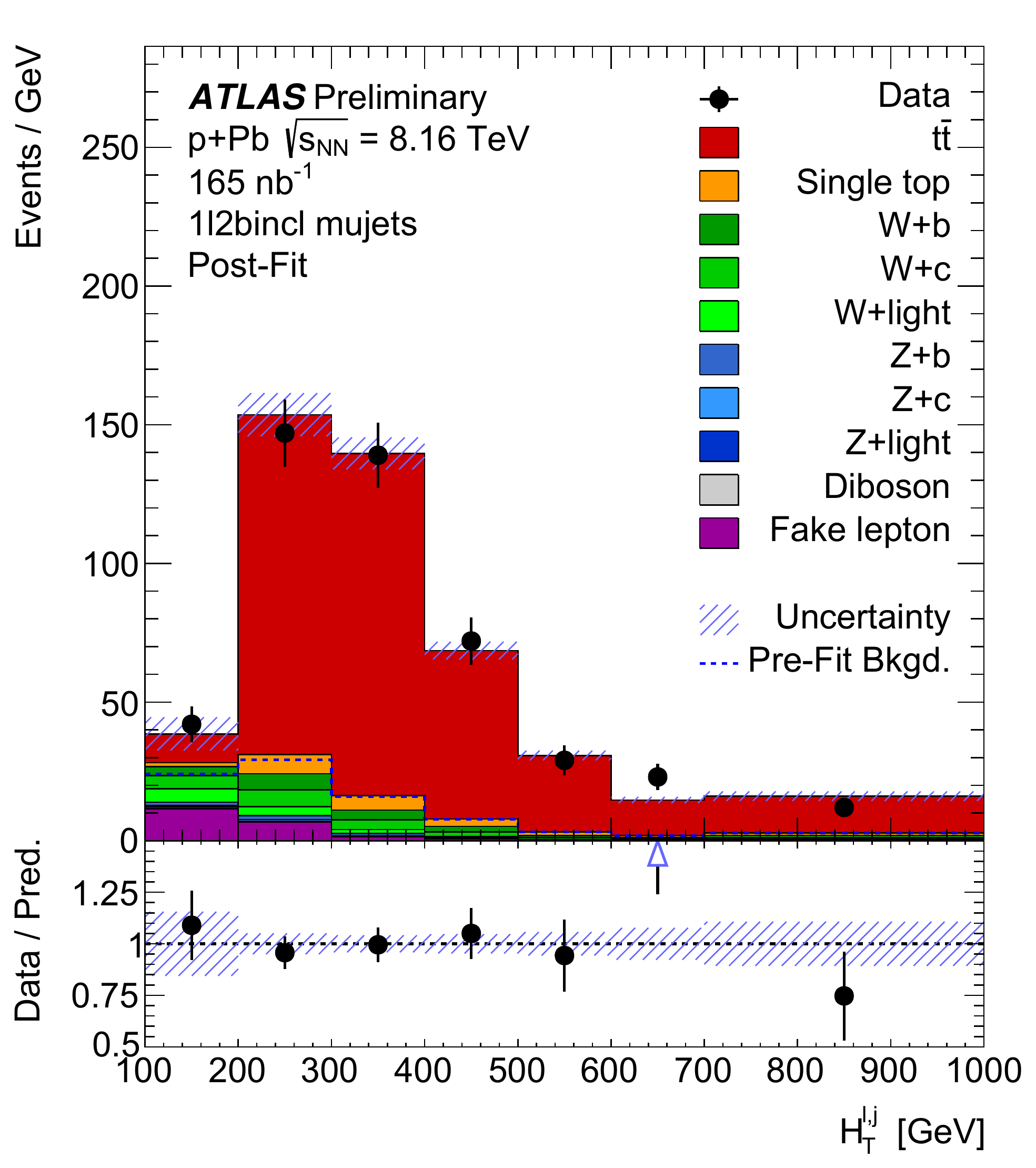}
\includegraphics[width=0.32\textwidth]{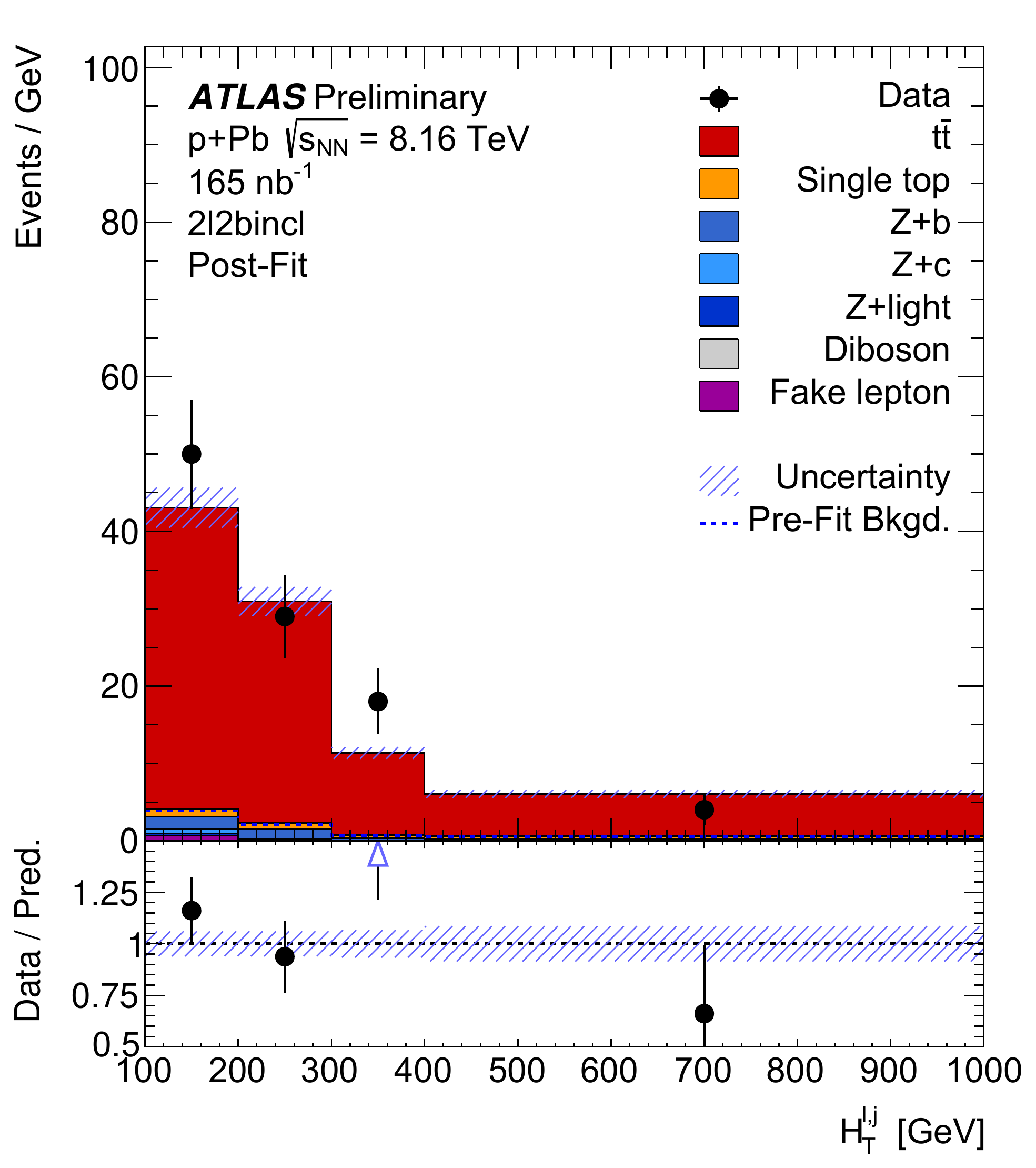}
\caption{Data post-fit plots representing the $\HTjl$ variable in the six SRs ($e+$jets: $1\ell1b$ and 
$1\ell2b\mathrm{incl}$, $\mu+$jets: $1\ell1b$ and $1\ell2b\mathrm{incl}$, dilepton: $2\ell1b$ and 
$2\ell2b\mathrm{incl}$.), with uncertainties represented by the hatched area. The full markers in the 
bottom panels show a ratio between data and a sum of predictions. Open triangles indicate bins with 
entries which are outside the ratio range.}
\label{fig:data_postfit_full}
\end{figure}

The signal strength parameter, denoted as \mutt{}, represents the observed signal yield in both the \ljets{} and dilepton final states, 
relative to the Standard Model (SM) expectation without nuclear parton distribution function (nPDF) effects. Its determination employs a profile-likelihood
 fit \cite{Cowan:2010js,Cranmer:1456844} to the \Ht{} data distributions across six Signal Regions, 
 where \Ht{} denotes the scalar sum of transverse momenta of leptons and HI jets. Figure in \ref{fig:data_postfit_full} depict the post-fit \Ht{} 
 distributions for six channels within the dilepton and \ljets{} SRs, distinguishing between exactly one and at least two $b$-tagged jets. 
 Observed distributions are in good agreement with those predicted by the fit.
  
%%%%%%%%%%%%%%%%%%%%%%%%%%%%%%
%%%%%%%%%%%%%%%%%%%%%%%%%%%%%%

The fit incorporates roughly 130 nuisance parameters that account for systematic uncertainties. These uncertainties are collectively treated as correlated 
across the Signal Regions. Primary contributors to the pre-fit uncertainties encompass fake-lepton and $W$+jets normalization in the \ljets{} channel, 
along with $Z$+jets normalization in the dilepton channel. The fit procedure mitigates the impact of fake-lepton uncertainties. 
The major factors contributing to the overall systematic uncertainty include signal modeling, jet energy scale, luminosity of the data sample, 
and fake-lepton background. A summary outlining grouped systematic uncertainties is presented in Table~\ref{tab:syst_group_data}. 
Overall, the total relative systematic uncertainty is estimated at 8\%.
%%%%%%%%%%%%%%%%%%%%%%%%%%%%%%
\begin{table}[htbp]
    \centering
    \caption{Breakdown of relative systematic uncertainties on the cross section in data. 
    Owing to rounding effects and small correlations between the different sources of uncertainties, 
    the total systematic uncertainty is different from the sum in quadrature of the individual sources.}
    \begin{tabular}{l|cc}
      \hline
      Source                                                                  &unc. up                &unc. down      \\%&symmetrized \\
      \hline
      Jet energy scale                                        & +0.048                &-0.044         \\%&0.043 \\
      $t\bar{t}$ generator                            & +0.048                &-0.043         \\%&0.043 \\
      Fake-lepton background                  & +0.030                &-0.027         \\%&0.025 \\
      Background                                                      & +0.030                &-0.025         \\%&0.027 \\
      Luminosity                                                      & +0.029                &-0.025         \\%&0.025 \\
      Muon systs.                                                     & +0.024                &-0.021         \\%&0.021 \\
      $W+$jets                                                        & +0.023                &-0.020         \\%&0.019 \\
      $b$-tagging                                                 & +0.022            &-0.021         \\%&0.020 \\
      Electron systs.                                         & +0.018                &-0.017         \\%&0.017 \\
      MC statistical uncertainties        & +0.011            &-0.010         \\%&0.010 \\
      Jet energy resolution                       & +0.005            &-0.004         \\%&0.001 \\
      $t\bar{t}$ PDF                                          & +0.001                &-0.001         \\%&0.002 \\
      \hline
      \hline
      Total syst.                                                     & +0.088                &-0.081         \\%&0.081 \\
      \hline
      \end{tabular}      
    \label{tab:syst_group_data}
\end{table}
%%%%%%%%%%%%%%%%%%%%%%%%%%%%%%

\section{Results}
The measured value of \mutt{} is used to derive the inclusive $t\bar{t}$ production cross section using the formula
\begin{equation}
  \sigma_{t\bar{t}}=  \mu_{t\bar{t}}\cdot A_\mathrm{Pb}\cdot \sigma^\mathrm{th}_{t\bar{t}}.
\end{equation}
Here, $\sigma^\mathrm{th}_{t\bar{t}}$ represents the predicted $t\bar{t}$ production cross section in nucleon-nucleon collisions, determined at NNLO precision, 
used as normalization for the signal $t\bar{t}$ samples in both \ljets{} and dilepton decay modes. The measured inclusive $t\bar{t}$ 
cross section for \pPb{} collisions is $\sigma_{t\bar{t}}= 57.9\pm 2.0~\mathrm{(stat.)\;^{+4.9}_{-4.5} ~\mathrm{(syst.)}}~\mathrm{nb}
= 57.9\;^{+5.3}_{-4.9} ~\mathrm{(tot.)}~\mathrm{nb}$. The total relative uncertainty stands at 9\%, mainly driven by systematic contributions.\par
\begin{figure}[!htb]
    \centering
    \includegraphics[width=0.75\textwidth]{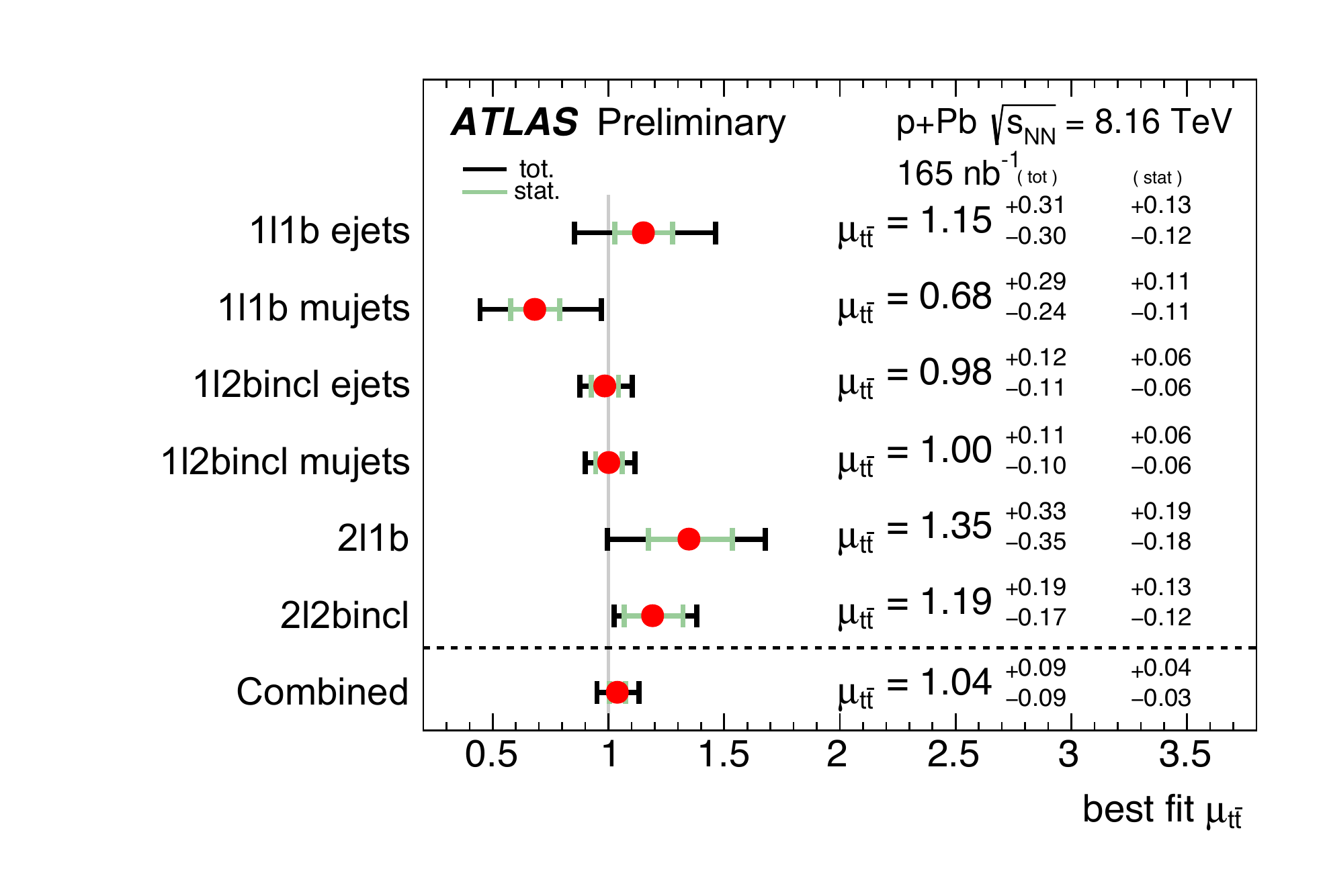} 
    \caption{Signal strengths with total and statistical uncertainties in different fit regions 
    and combined.}
    \label{fig:mu}
\end{figure}

The background-only hypothesis is dismissed with a significance exceeding 5 standard deviations, confirming the observation of the $t\bar{t}$ 
process in \pPb{} collisions by ATLAS. Figure~\ref{fig:mu} showcases the signal strength \mutt{} 
obtained separately in each region and through the combined fit. The two \ljets{} channels with the highest signal yield (in the $\geq 2b$ region) 
produce very close results. Slight variations are observed in the signal strength between the dilepton channels, with a slightly higher \mutt{} 
preferred in the dilepton channels compared to the \mujets{} $1b$ channel, but consistent within uncertainties.
The precision of the \mutt{} value is limited by systematic uncertainties in the \ljets{} SRs, while statistical uncertainties dominate 
in the dilepton SRs. Additionally, separate fits of \mutt{} to the combined four \ljets{} and two dilepton SRs both exceed $5\sigma$, confirming the observation 
of $t\bar{t}$ production also in the individual \ljets{} and dilepton channels. This marks the first report of the latter in \pPb{} collisions at the LHC.

%%%%%%%%%%%%%%%%%%%%%%%%%%%%%%
%%%%%%%%%%%%%%%%%%%%%%%%%%%%%%

\begin{figure}[!htb]
  \centering
  \includegraphics[width=0.75\textwidth]{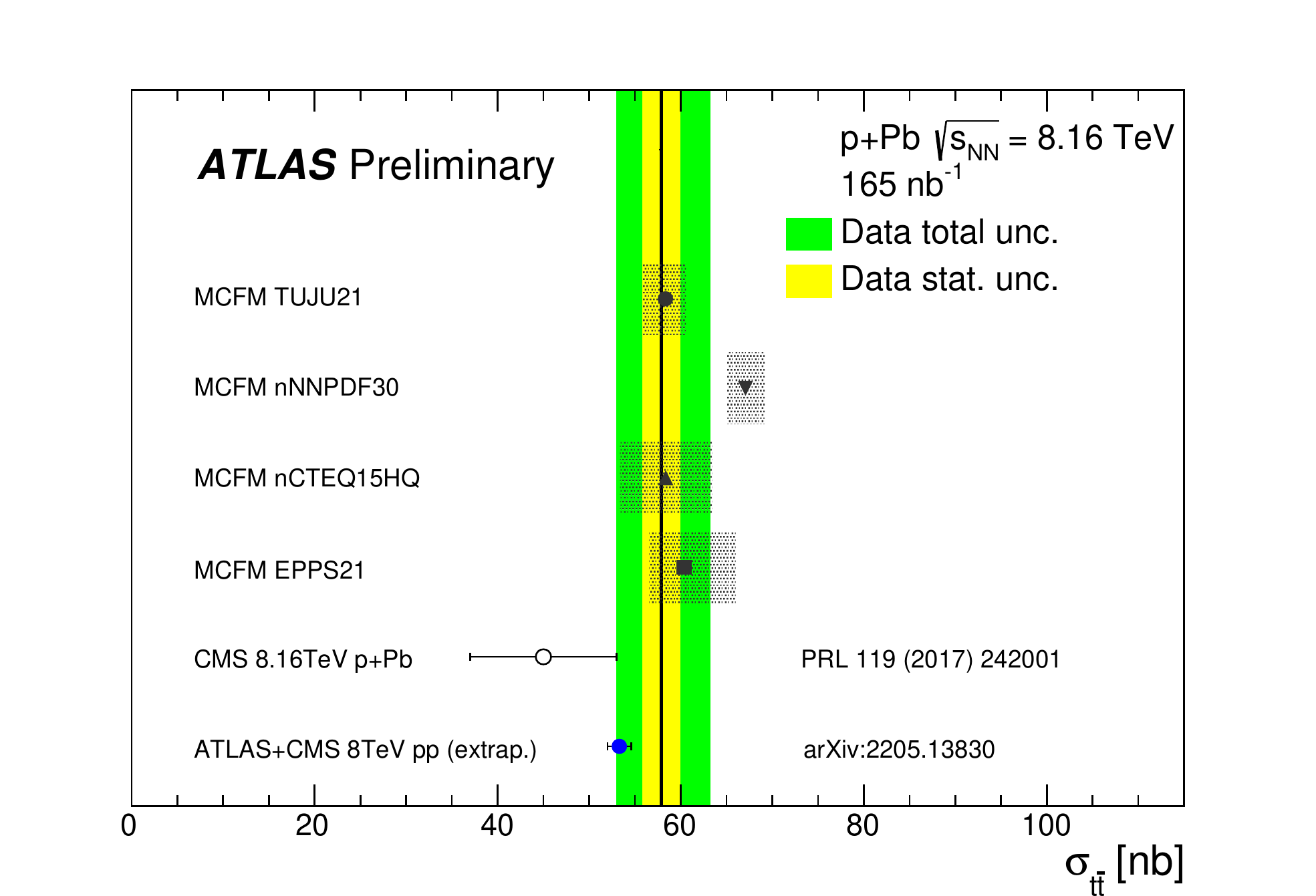}\\
  \caption{Comparison between observed and predicted values of \xsectt{} as well as with the CMS Collaboration
  measurement of \xsectt{} in \pPb\ collisions at $\sqrt{s_\mathrm{NN}}=8.16~\mathrm{TeV}$~\cite{CMS:2017hnw},
  and the combined measurement of $t\bar{t}$ production cross section in $pp$ collisions at 
  $\sqrt{s}=8~\mathrm{TeV}$ from ATLAS and CMS collaborations~\cite{ATLAS:2022aof}. 
  The latter is extrapolated to the center-of-mass energy of this measurement and 
  also using the $A_\mathrm{Pb}$ factor.
  Predictions are calculated at NNLO precision using the \mcfm\ code~\cite{Campbell:2016dks} scaled
  to the \pPb\ system and given for different nPDF sets. 
  The uncertainty on predictions represents the uncertainty on internal PDF.
  The solid black line indicates the measured value.
  The combined statistical and systematic uncertainty of the measurement is shown in green while 
  the statistical component is depicted in yellow.
  %while the statistical uncertainty is represented by the yellow error band. 
  }
  \label{fig:xsec}
\end{figure}

The combined measurement of the $t\bar{t}$ production cross section \xsectt{} showed in Figure~\ref{fig:xsec} is compared with the corresponding result 
obtained by CMS~\cite{CMS:2017hnw}, both acquired using \pPb\ collisions at $\sqrt{s_\mathrm{NN}}=8.16~\mathrm{TeV}$. These two measurements exhibit agreement within 
the range of uncertainties. Additionally, the most precise $t\bar{t}$ production cross section measured in $pp$ collisions at $\sqrt{s}=8~\mathrm{TeV}$, derived 
from a combination of ATLAS and CMS data~\cite{ATLAS:2022aof}, is presented. This cross-section value is extrapolated to the center-of-mass energy of this 
particular measurement using predictions from \toppp~v2 and scaled by $A_\mathrm{Pb}$ to the \pPb\ system. The extrapolated cross section bears a 2.5\% 
relative uncertainty and remains independent of nPDF dependencies. The measured cross section is compared with NLO calculations obtained using 
the \mcfm\ generator~\cite{Campbell:2016dks}, scaled to NNLO precision in QCD using a k-factor (k=1.139) derived from the \toppp\ generator. The \mcfm\ 
calculations utilize four distinct nPDF sets: EPPS21~\cite{Eskola:2021mjl}, nCTEQ15HQ~\cite{Kusina:2020lyz}, nNNPDF30~\cite{Ubiali:2014bva}, 
and TUJU21~\cite{Helenius:2022rmf}. The nNNPDF30 nPDF set exhibits the largest discrepancy when compared with the measured cross-section value. 
In contrast, the remaining nPDF sets showcase good alignment with the measured cross-section value.

%%%%%%%%%%%%%%%%%%%%%%%%%%%%%%
%%%%%%%%%%%%%%%%%%%%%%%%%%%%%%

%\FloatBarrier
%Summary
\section{Conclusion}

This proceeding presented measurement of top-quark pair production in \pPb\ collisions at a center-of-mass energy of $\sqrt{s_\mathrm{NN}}=8.16~\mathrm{TeV}$ 
per nucleon pair, conducted with the ATLAS experiment. The observation of top-quark pairs occurs through the distinct \ljets{} and dilepton channels, 
with final states containing electrons and muons. The observation of top-quark pair production in the dilepton channel surpasses 8 standard deviations, 
marking its first observation in the \pPb\ system at the LHC. The total integrated cross section is measured with a total relative uncertainty of 9\%, 
establishing this as the most precise determination of the $t\bar{t}$ cross section in nuclear collisions to date. While the precision in the individual 
$t\bar{t}$ decay channels is constrained by systematic uncertainties in the \ljets\ channel, the dilepton region is predominantly limited by statistical uncertainties.

The measured cross section is compared with the CMS measurement in the \pPb\ system and the combined measurement from ATLAS and CMS in $pp$ collisions at 
$\sqrt{s}=8~\mathrm{TeV}$. The latter is appropriately scaled to the \pPb\ system and the energy $\sqrt{s}=8.16~\mathrm{TeV}$ of this analysis. Additionally, the measured cross section 
is compared against NNLO calculations derived from various nPDF sets. The measured 
cross section aligns closely with previous measurements and Standard Model predictions. With its enhanced precision, this measurement promises to offer 
constraints on nPDFs, particularly in the high-$x$ region.

\section*{Acknowledgement}
The author gratefully acknowledge the support from the project IGA\_PrF\_2023\_005 of Palacky University.

%-------------------------------------------------------------------------------
% Auxiliary material - comment out the following line if you do not have any.
%\include{ANA-TOPQ-2018-43-PAPER-auxmat}
%-------------------------------------------------------------------------------

%-------------------------------------------------------------------------------
% Extra tables etc. for HepData - comment in the following line if you have any.
% \include{ANA-TOPQ-2018-43-PAPER-hepdata}
%-------------------------------------------------------------------------------

\end{document}